%% file: microscan_miri.tex
\documentclass[10pt]{spie}  
\usepackage[]{graphicx}
\usepackage{rotating}
\usepackage{times}
\usepackage{multirow}

        \input alphabet.tex     
        \input abrege.tex        
      \input abrmath.tex       
      
\title{Optical performance of the JWST/MIRI flight model: characterization of the point spread function at high-resolution} 

\author{Guillard P.\supit{a,b}, Rodet T.\supit{c}, Ronayette S.\supit{d}, Amiaux J.\supit{d}, Abergel A.\supit{b}, Moreau V.\supit{d},  Augueres, J.L.\supit{d}, Bensalem, A.\supit{d}, Orduna, T.\supit{d}, Nehm\'e, C.\supit{d}, Belu, A.~R.\supit{d}, Pantin, E.\supit{d}, Lagage, P.-O.\supit{d} Longval, Y.\supit{b}, Glasse, A.~C.~H.\supit{e}, Bouchet, P.\supit{d}, Cavarroc, C.\supit{b}, Dubreuil, D.\supit{d}, Kendrew, S.\supit{f}
\skiplinehalf
\supit{a}Spitzer Science Center, IPAC, California Institute of Technology, Pasadena, CA-91125, USA \\
\supit{b}Institut d'Astrophysique Spatiale, CNRS/Universit\'e Paris Sud 11, Orsay, France \\
\supit{c}Laboratoire des Signaux et Syst\`emes/CNRS/Universit\'e Paris Sud 11, Gif-sur-Yvette, France \\
\supit{d}Laboratoire d'Astrophysique, Instrumentation-Mod\'lisation, de Paris-Saclay (CEA/Irfu, Universit\'e Paris-Diderot, CNRS/INSU), CEA Saclay, Gif-sur-Yvette, France. \\
\supit{e}UK ATC, Royal Observatory, Blackford Hill, Edinburgh, EH9 3HJ, Scotland, UK \\
\supit{f}Leiden University, Netherlands \\
}

\authorinfo{Further author information: (Send correspondence to P. Guillard)\\E-mail: guillard@ipac.caltech.edu}

\begin{document} 
  \maketitle 

\begin{abstract}
The Mid Infra Red Instrument (MIRI) is one of the four instruments onboard the James Webb Space Telescope (JWST), providing imaging, coronagraphy and spectroscopy over the $5-28\,\mu$m band. To verify the optical performance of the instrument, extensive tests were performed at CEA on the flight model (FM) of the  Mid-InfraRed IMager (MIRIM) at cryogenic temperatures and in the infrared. This paper reports on the point spread function (PSF) measurements at 5.6$\,\mu$m, the shortest operating wavelength for imaging. At 5.6$\,\mu$m, the PSF is not Nyquist-sampled, so we use am original technique that combines a microscanning measurement strategy with a deconvolution algorithm to obtain an over-resolved  MIRIM PSF. The microscanning consists in a sub-pixel scan of a point source on the focal plane. A data inversion method is used to reconstruct PSF images that are over-resolved by a factor of 7 compared to the native resolution of MIRI. We show that the FWHM of the high-resolution PSFs were $5-10$~\% wider than that obtained with Zemax simulations. The main cause was identified as an out-of-specification tilt of the M4 mirror. After correction, two additional test campaigns were carried out, and we show that the shape of the PSF is conform to expectations. The FWHM of the PSFs are $0.18-0.20$~arcsec, in agreement with simulations. $56.1-59.2$\% of the total encircled energy (normalized to a 5~arcsec radius) is contained within the first dark Airy ring, over the whole field of view. At longer wavelengths ($7.7-25.5\,\mu$m), this percentage is $57-68\,$\%. MIRIM is thus compliant with the optical quality requirements. This characterization of the MIRIM PSF, as well as the deconvolution method presented here, are of particular importance, not only for the verification of the optical quality and the MIRI calibration, but also for scientific applications.   
\end{abstract}



\section{Introduction}
\label{sec:intro}  

\begin{figure}
  \begin{center}
    \includegraphics[width=\textwidth]{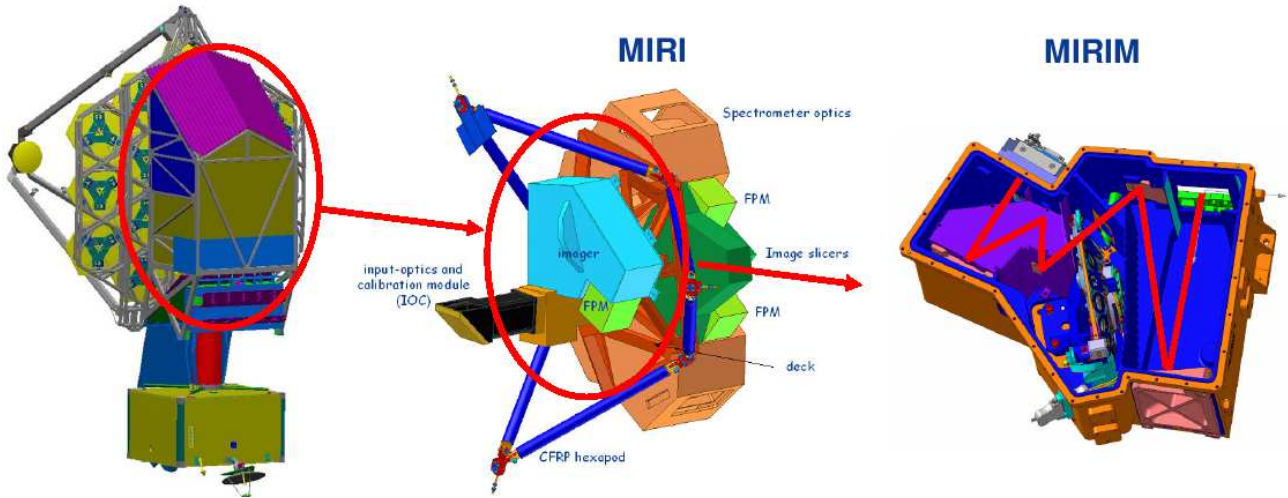}
    \caption{\textit{Left:} The Mid Infra Red Instrument (MIRI) as part of the scientific payload of the JWST.  \textit{Middle:} The MIRI Optical Bench Assembly (MIRI-OBA), with the imager in blue. \textit{Right:} The MIRIM (Mid Infra Red IMager) mechanical layout, with the filter wheel assembly (FWA). MIRIM shall provide the following science functions: photometric Imaging between 5 and 27$\, \mu$m, coronagraphy between 10 and 27$\, \mu$m, low resolution spectroscopy between 5 and 10$\, \mu$m.}
    \label{fig_MIRI_payload}
  \end{center}
\end{figure}

The JWST Mid Infra Red Instrument (MIRI, Wright et al. 2004\cite{Wright2004}, Rieke et al. 2005\cite{Rieke2005}) is part of the scientific
payload of the James Webb Space Telescope (JWST) (Fig.~\ref{fig_MIRI_payload}).  Three models were built before the Flight Model (FM): the  
Structural Qualification Model (SQM), the verification model (VM), and the Engineering and Test Model (ETM).
The different steps leading to the FM integration of the Mid-Infra Red IMager Optical Bench (MIRIM-OB), and the principal results associated to the SQM for vibration, and the ETM for optical performance, are presented in Amiaux et al. 2008\cite{Amiaux2008}. 

To verify the optical performance of the MIRIM FM, extensive tests were performed from Dec. 2008 to Dec. 2009 at CEA, in the infrared and at cryogenic temperatures. 
An overview of the FM tests is presented in Ronayette et al. 2010\cite{Ronayette2010}. In this paper we focus on the characterization of the point spread function (PSF) of MIRIM, and particularly at $5.6\,\mu$m. This is the shortest operating wavelength, and therefore the most critical to assess the optical quality of the imager. We use a microscanning technique to measure and reconstruct the PSF at a higher resolution than the native one of the instrument. This allows us to characterize the MIRIM PSF very accurately and verify that the imager is compliant with PSF requirements. This study will be particularly helpful for the photometric calibration of MIRI, but also for scientific applications.

We present three series of tests, resulting from the FM1 (December 2008), FM2 (April 2009) and FM4 (December 2009) cryogenic test campaigns at CEA. The field of view (henceforth FoV) and PSF measurements carried out during the FM1 campaign showed a defect in the optical quality (vigneting issue and wider PSF than expected, see Sect.~\ref{subsec:test_campaigns}). We then show the results of the FM2 and FM4 campaigns, after correction of this defect.
This paper is organized as follows: Sect.~\ref{sec:instru_setup} briefly presents the instrumental setup used to simulate the JWST beam and perform the cold tests. After a short description of the data reduction (Sect.~\ref{sec:data_reduction}),  we derive  the response curve of the detector and show how we correct the images for the non-linearity of the detector (Sect.~\ref{sec:linearity}). Then, Sect.~\ref{sec:microscan} describes the microscanning test and the method used to reconstruct the MIRIM PSF  at high resolution. Sect.~\ref{sec:results} presents our results about the characteristics of the PSF. We finally present our conclusions in Sect.~\ref{sec:conclusion}.

\section{Instrumental setup: the MIRIM test bench at CEA}
\label{sec:instru_setup}

\begin{figure}
  \begin{center}
    \includegraphics[width=0.65\textwidth]{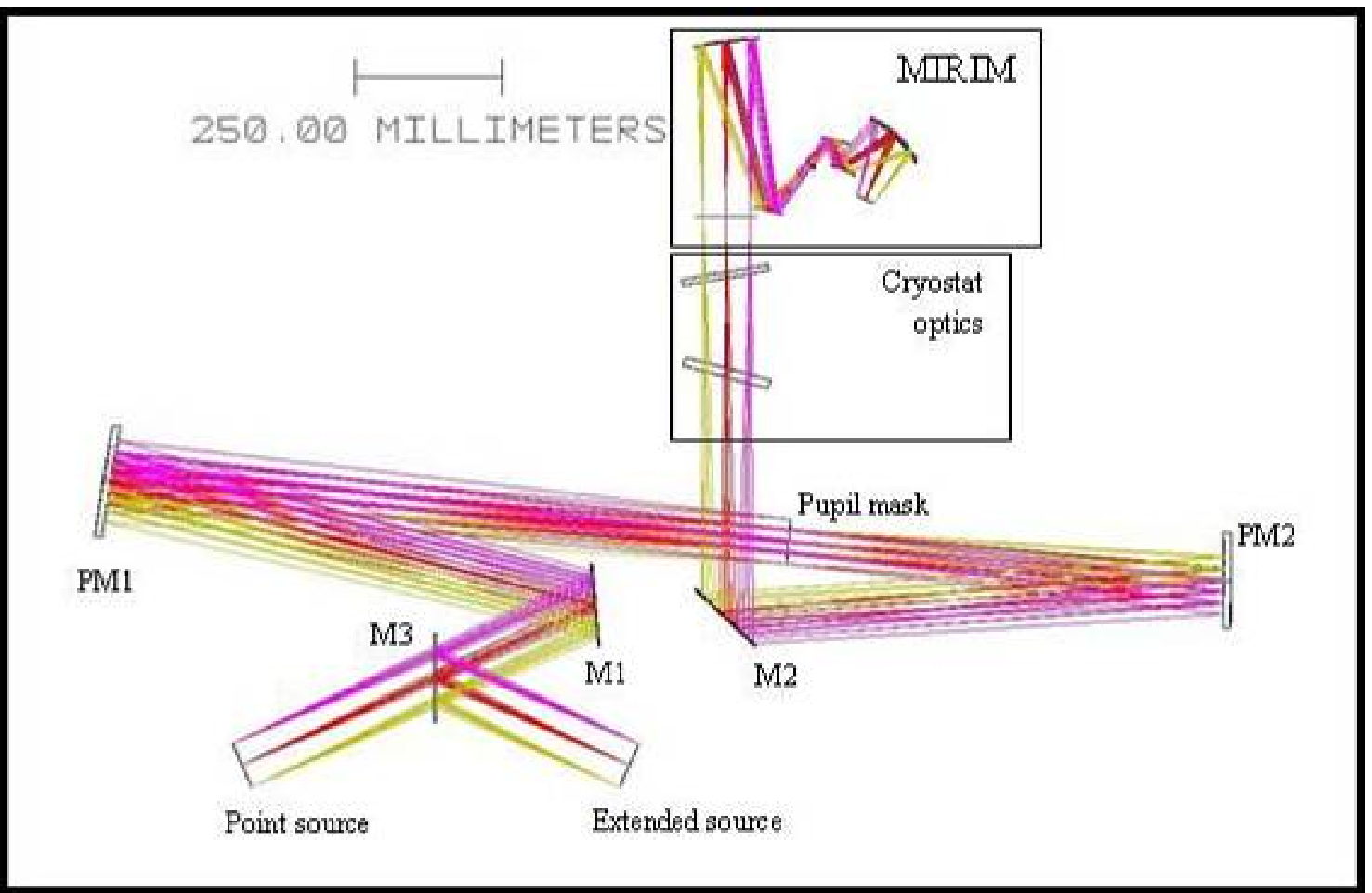}
    \includegraphics[width=0.34\textwidth]{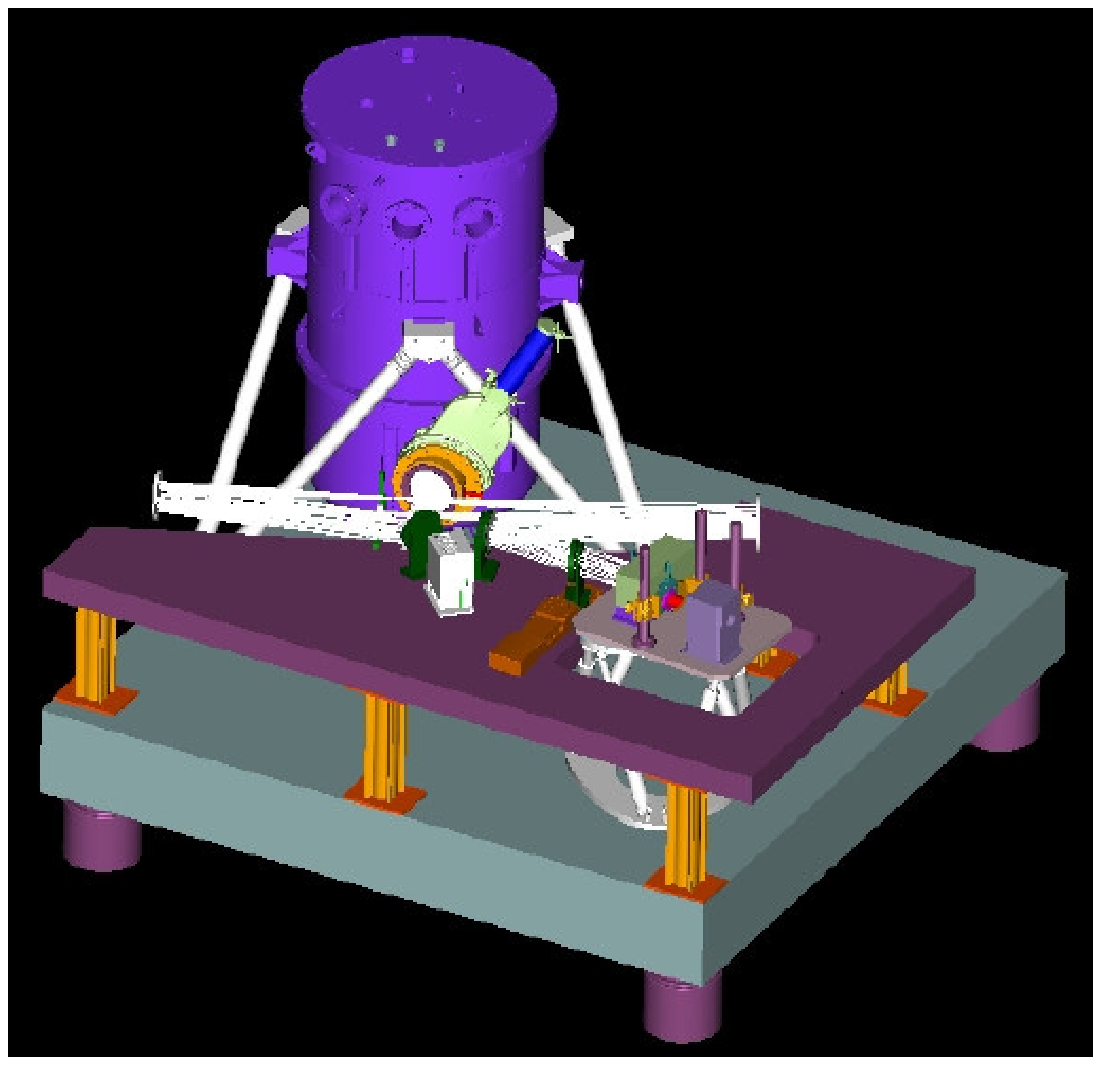}
    \caption{The MIRIM test bench. The left panel shows the optical scheme of the Telescope Simulator (TS) that simulates the JWST beam and pupil to characterize MIRIM in image or coronographic mode. The warm TS is installed outside the cryostat (in blue on the right panel) that contains MIRIM. The TS allows us to use  a point or extended source.}
    \label{fig_testbed_simu_telescope}
  \end{center}
\end{figure}

The verification of the MIRIM optical quality has been done with an ambient temperature Telescope Simulator (TS), installed outside the  helium-cooled cryostat that contains MIRIM-OB (Fig.~\ref{fig_testbed_simu_telescope}). The TS simulates the optical beam delivered by the JWST. The optical conception of the TS is based on two off-axis parabolic mirrors (PM1 and PM2) and a pupil mask in between. The pupil mask is the pupil of the telescope simulator (so-called the ``STOP'') and it reproduces the pupil of JWST. It is mounted on a motorised translation and rotation stages (4 axis: 3 translations and one rotation), which allows us to co-align the telescope simulator and the MIRIM pupils. The FoV is $80\, \rm mm \times 80\, mm$ for a MIRIM FoV of $72\, \rm mm \times 72\, mm$. An analysis of the image quality shows that the distortion amount of the TS is 0.33\% for the useful field (Y. Longval, private communication). The aperture is $F/20$ in the image plane. The exit pupil location is 3017.5~mm from the image plane for a diameter of 151.6~mm. The  pupil is 75.4 mm in diameter. To implement the MIRIM cryostat in a minimum space, folding mirrors are used.

An IR point source with a shutter is mounted on a remotely controlled hexapod, that has 6 degrees of freedom ($X$, $\theta _X$, $Y$ , $\theta _Y$, $Z$, $\theta _Z$), with an accuracy of $1\,\mu$m. The minimum diameter of the point source is 30$\, \mu$m, and the temperature of the source can be adjusted at 1150~K or 2000~K.
An extended black body source (400~K) can also be used for flat fields for instance.
The 10K screen in the cryostat is fitted with a neutral density (ND) that reduces the flux due to the ambiant warm temperature (295~K) background. 

\section{Data acquisition settings and reduction}
\label{sec:data_reduction}

For the microscanning, the shutter is used in chopping mode. Several frames are taken for each cycle (open$-$close shutter). For each cycle, the first frame is skipped because during this frame the shutter is closing. Then, the following 3 frames are skipped because of latency effects on the detector since the point source is very bright.  For the FM2 test campaign, sub-arrays of $72 \times 64$ pixels were read out to speed up the data acquisition and reduction, with a detector integration time $DIT = 0.86$~s. Each exposure  comprised  10 cycles, and the final image is obtained by taking a median from these 10 cycles, corresponding to a total integration time of $T_{\rm exp} = 132.3$~s. 
For the FM4 campaign, the acquisition settings were different.  $136 \times 128$ sub-arrays were read out, with a detector integration time $DIT = 0.30$~s.  The finale image is a median created from 50 cycles, corresponding to a total integration time of $T_{\rm exp} = 183.3$~s. The final images are also flat-fielded.

\section{Linearity correction for the response of the detector}
\label{sec:linearity}

The MIRIM detector (a Raytheon SB305 engineering model type) is a $1024 \times 1024$ pixels SiAs array, loaned to CEA by JPL for the purpose of the test campaigns. The detector configuration is a Sensor Chip Assembly (SCA), i.e. an ultra-light mechanical and electrical housing. CEA has developed an appropriate housing (similar to the MIRI focal plane) to mount the IR detector onto the imager for the cold tests. Thus, qualitative analysis of the optical properties, including stray light performances, are possible. Note that the detector and the readout electronics used for these tests were not FM hardware.

\begin{figure}
      \includegraphics[height=7.5cm, angle = 90]{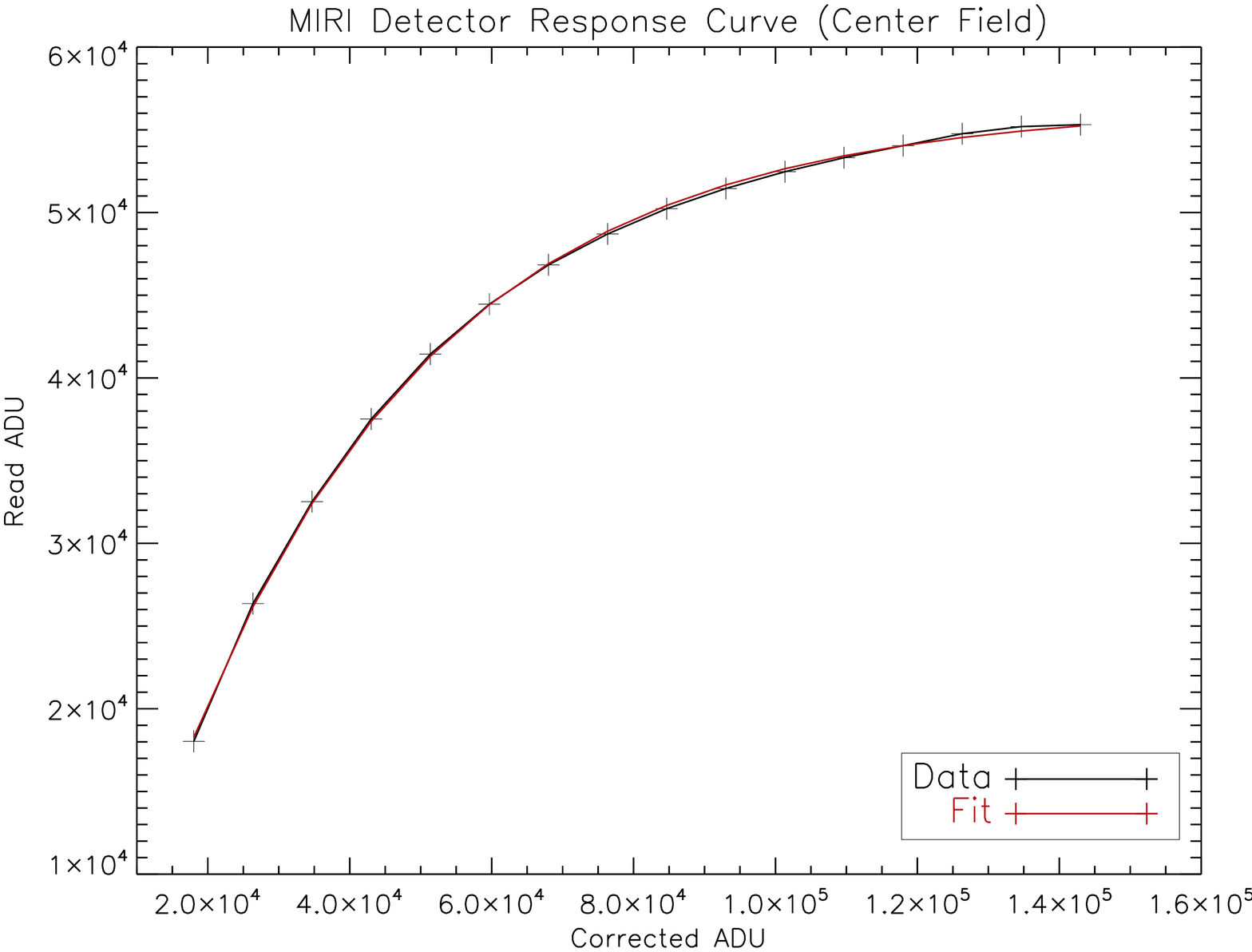}
      \includegraphics[height=9.9cm, angle = 90]{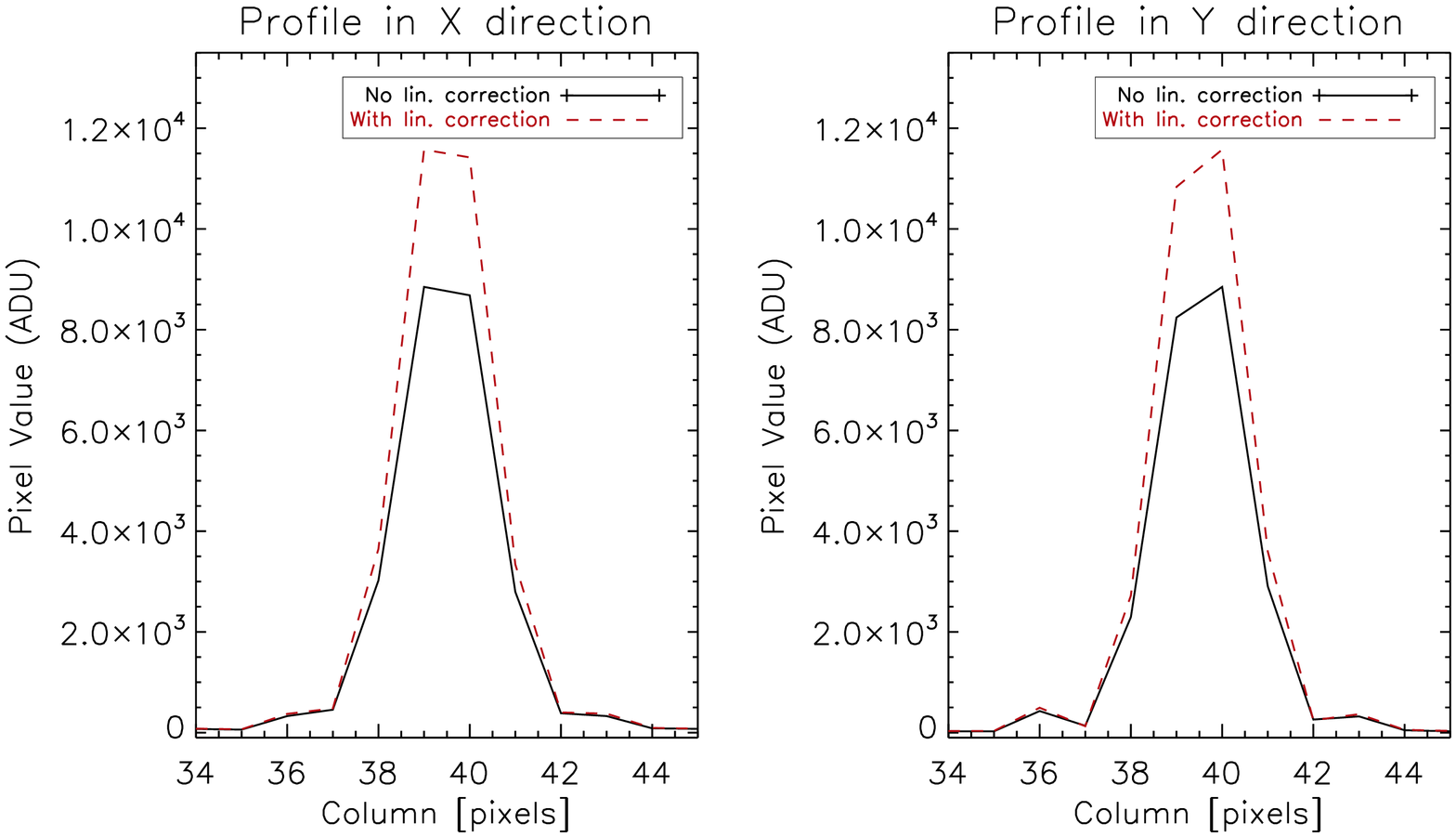}
    \caption{\textit{Left:} response curve of the MIRI SB305 detector obtained at CEA. The back curve show the average read ADU values of the pixels in the center of the field as a function of the corrected values (see text for details). The red curve indicates an exponential fit (see \S~\ref{sec:linearity}, Eq.~\ref{eq:response}). \textit{Right:} impact of the linearity correction on the low-resolution PSFs. The plots show MIRIM PSFs profiles at $5.6\, \mu$m in two perpendicular directions. The pixel size is $25\, \mu$m. The black line represents data that have not been corrected for the detector response. The red dashed line shows the same PSF profile when the linearity correction is applied (see Sect.~\ref{sec:linearity}). The PSF shown here were taken during the FM2 campaign, in the center of the MIRIM field of view.}
    \label{fig_MIRI_Response_Curve}
 %
    \label{fig_comp_lin_correction}
\end{figure}

We tested the response of the SB~305 detector at CEA, by measuring the flux of an extended source with increasing integration times. The temperatures of the detector and the extended source are constant (5~K and 30~K, respectively), the integration time is the only parameter that changes.  No filter has been used. Fig.~\ref{fig_MIRI_Response_Curve} presents the curve of the detector response, i.e. the read value of the flux (in ADU) as a function of the ``corrected value''. The corrected value is the expected flux if the response were linear. The 30~K blackbody source does not allow us to explore the linear part of the response curve, at low fluxes. With the smallest integration time (2.7~s for a full frame), we reach a signal of $26\,360$~ADU, which is already in the non-linear regime (Fig.~\ref{fig_MIRI_Response_Curve}). To measure low fluxes in the linear regime, we used colder black body sources. An accurate sampling of the response curve at low ADU values  shows that the response is linear for ADU values lower than $\approx 16\,000$ (V. Moreau, private communication) and follows the law:
\begin{eqnarray}
\label{eq:signal-linear-MIRI}
S_{\rm lin} = \alpha \times t_{\rm int} + \beta \ \ ,
\end{eqnarray}
where $S_{\rm lin}$ is the signal in ADU in the linear part of the response, $\alpha = 3067$~ADU~s$^{-1}$ is the slope of the linear response and $\beta = 18030$~ADU the offset. Note that absolute photometric measurements were not possible because the quantum efficiency and conversion gain of the detector used for these tests were not characterized. 

Fig.~\ref{fig_MIRI_Response_Curve} shows that the detector is non linear over a wide range of ADU values.  
The response curve can fitted with an exponential function:
\begin{equation}\label{eq:response}
S_{\rm read} = A - B \, e^{-S_{\rm cor} / C} \ ,
\end{equation}
where $S_{\rm read }$ is the read signal in ADU and $S_{\rm cor}$ the corrected one. We found $A = 56\,418.5$, $B = 62\,939.1$ and $C = 35\, 990.5$. 

To apply the linearity correction, we derive $S_{\rm cor}$ from $S_{\rm read }$ and correct all the individual images. 
To quantify the impact of the non-linear response curve on the PSF measurements, we provide two sets of test data: one that is not corrected for the detector response, and the other which is corrected for the non-linearity. We perform the PSF analysis on the two data sets and we compare the results. 
Fig.~\ref{fig_comp_lin_correction} compares the PSF profiles when the linearity correction is applied or not. 
The correction for the non-linear response of the detector improves the sharpness of the MIRIM PSFs because the deviation to the linearity is greater at high fluxes than the one at low signal. The FWHM of the corrected PSFs are $\sim 5\,$\% lower than the raw ones. Note that, although we do see a non-linear response in the FM detectors, the deviation to a linear behaviour is not as severe as in the detector used for these tests.

\section{Microscanning test and data inversion method}
\label{sec:microscan}

We describe here the aims of the microscanning tests and the deconvolution method we used to reconstruct over-resolved images of the PSF. The results are discussed in Sect.~\ref{sec:results}. 

\subsection{Aims and method}

The FWHM of the MIRIM PSF at $5.6\,\mu$m is 0.185'', i.e. less than two pixels (1 pixel\footnote{The physical size of a pixel on the detector is $25\, \mu$m}$ = 0.11''$), so the PSF is not Nyquist-sampled at this wavelength. The goal of the microscanning test is to obtain an over-resolved image of the PSF from observed multiple low-resolution (LR) images. Such a resolution enhancement approach has been an active research area (e.g. Park et al. 2003\cite{Park2003}, Molina et al. 2006\cite{Molina2006}, Idier 2008\cite{Idier2008}), and it is sometimes called over-resolution image reconstruction.

\begin{figure}
  \begin{center}
    \includegraphics[width=0.35\textwidth]{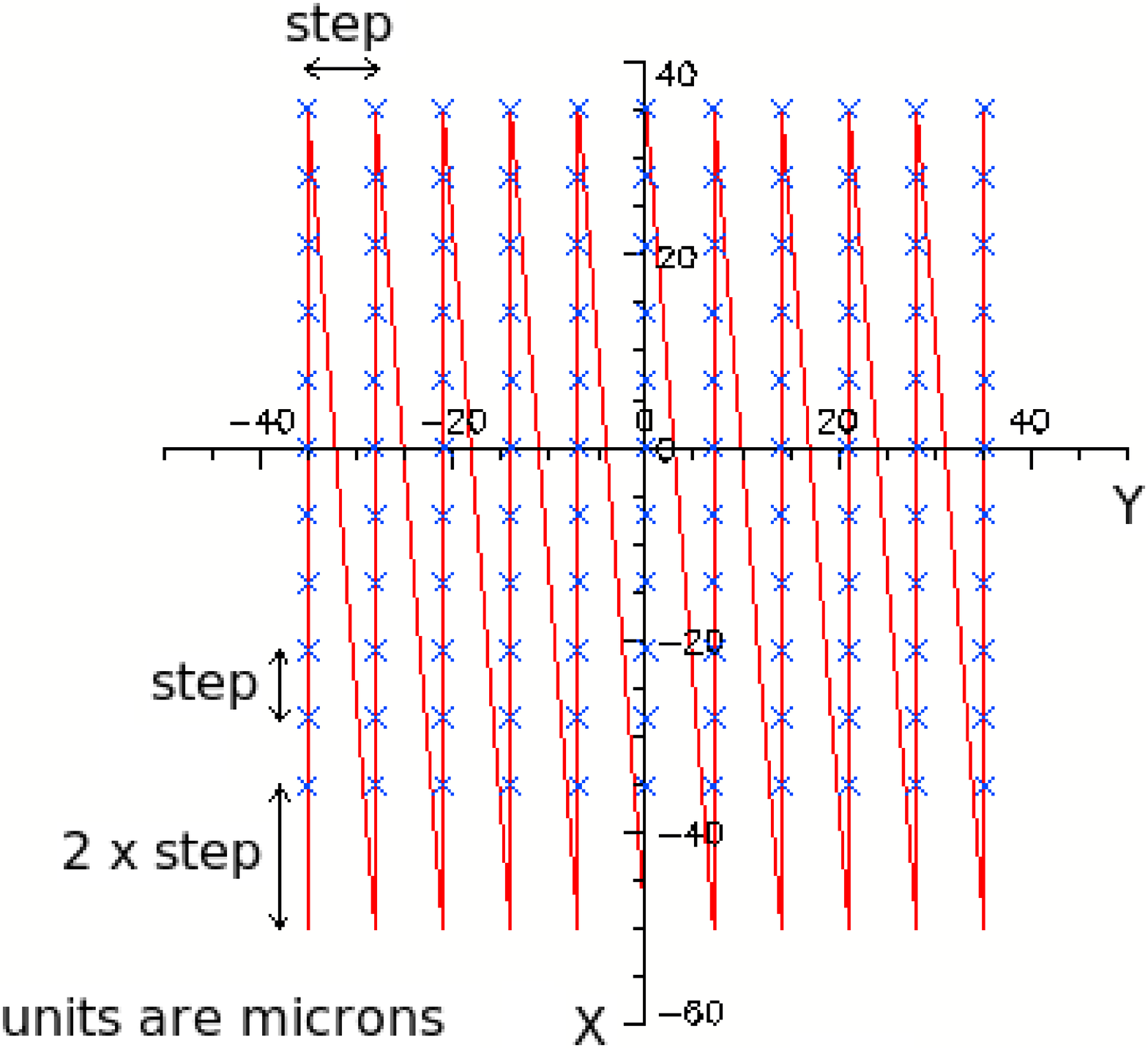}
     \includegraphics[width=0.64\textwidth]{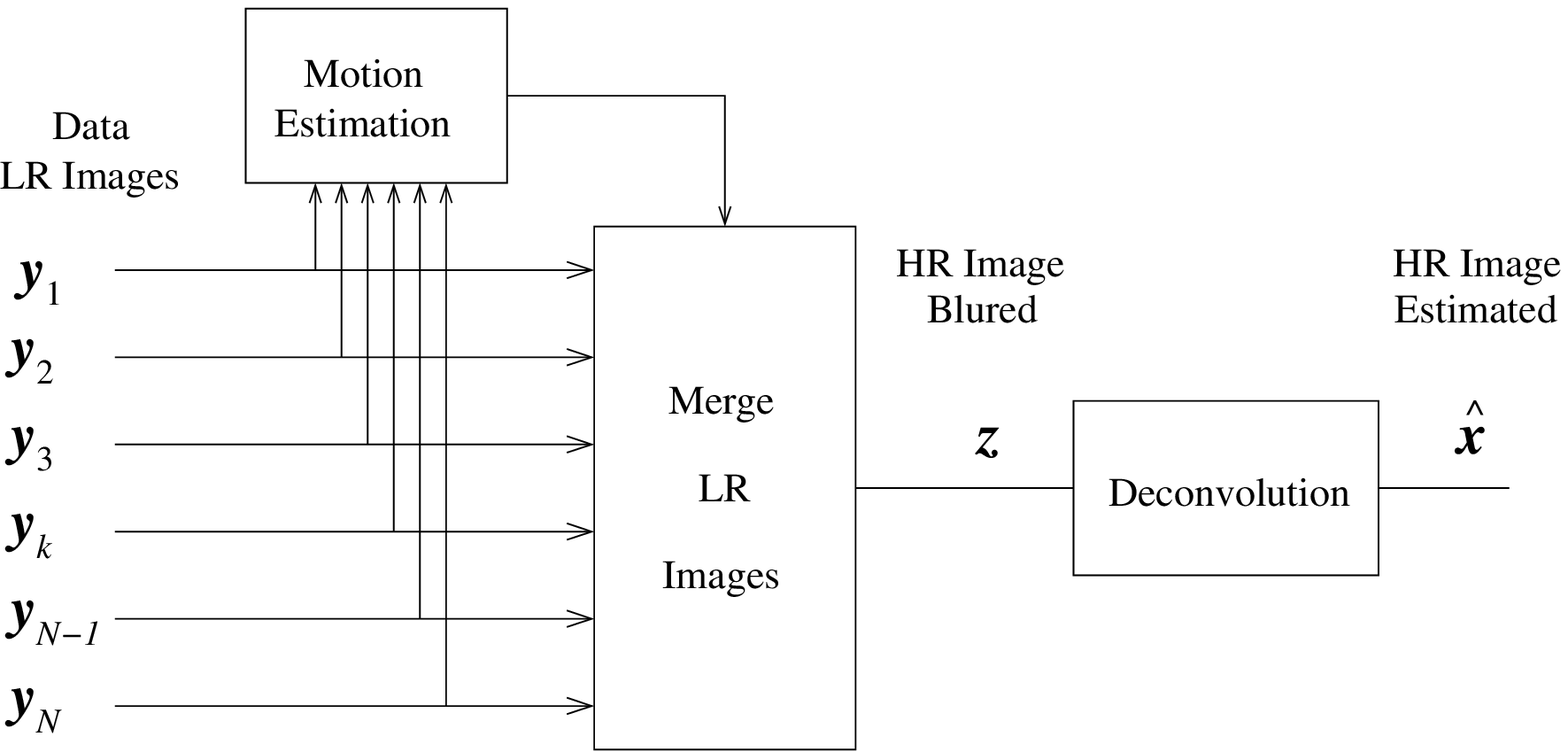}
    \caption{\textit{Left:} Pattern of the microscan. A $11 \times 11$ points scanning grid is performed, corresponding to a total scanning area of 1 pixel on the detector plane. The source is moved by steps of $7$~$\mu$m on the hexapod mount of the telescope simulator. This corresponds to individual displacements of  $2.5$~$\mu$m on the detector plane. \textit{Right:} Sketch of the different steps to reconstruct the high-resolution image (inverse problem).}
    \label{fig_schema_scan}
    \label{fig_inverseModel}
  \end{center}
\end{figure}

The method consists in scanning a point source in a fine spatial resolution.  The microscan pattern is given in Fig.~\ref{fig_schema_scan}. A $11 \times 11$ points grid was done, corresponding to a total scanning area of 1 MIRIM pixel on the detector plane, i.e., an area of  $25^2$~$\mu$m$^2$. The source has been moved by steps of $7$~$\mu$m on the hexapod mount. This corresponds to individual displacements of  $2.5$~$\mu$m on the detector plane. The best focus was searched at the beginning of each test serie. Microscans were performed at 9 positions of the imager FoV.

We summarize how we reconstruct the over-resolved images. Fig.~\ref{fig_inverseModel} shows a sketch of the inverse problem.  Let us denote $\xb$ the high-resolution (HR) PSF. 
The aim of the deconvolution method is to deduce $\xb$ from $N$ low-resolution images $\yb_k$ (the observed data). Assuming that we start from the high-resolution image $\xb$, the \textit{forward} problem can be expressed as
\begin{equation}
\label{eq:directproblem}
  \yb_k =  \Sb \Rb \Tb_k\xb + \nb_k
\end{equation}
where $\Tb_k$, $\Rb$ et $\Sb$ are three matrices associated with the operations of translation, convolution by the impulse-response of the detector (PSF), and down-sampling, respectively. $k\in \{1,\ldots,N\}$, where $N$ is the total number of images taken at different positions on the detector. $\nb$ is the noise, whose level is assumed to be commensurate for all the images. Note that in Eq.~\ref{eq:directproblem}, the only operation that is different from an image to another is the translation $\Tb_k$.


The first step to reconstruct HR images is to co-add the individual LR images on a fine grid, over-sampled by a factor $f$. To do so, the relative translations between the LR images have to be known with precision. Unfortunately, the mechanical control of the hexapod (where the point source is mounted) is not perfect, which results in mismatches between the expected position and the effective observed location of the source on the detector.
Therefore we chose to estimate the shifts between images by cross-correlation. 
The three steps of the reconstruction are the following:
\begin{enumerate}
\item we first estimate the translation between the low-resolution (LR) images by cross correlation (see Sect.~\ref{subsec:translations}). 
\item then, we co-add the LR images on a fine grid, assuming that there is no convolution by the PSF of the detector.
\item finally, we solve the deconvolution problem with a Bayesian approach to estimate $\xb$ (see Sect.~\ref{subsec:deconvolution}).
\end{enumerate}
The deconvolution algorithm is a simplified version of a more complex algorithm developped by Rodet et al. 2009\cite{Rodet2009} to perform over-resolution with spectral mapping data provided by the  \textit{Spitzer} telescope InfraRed Spectrometer (IRS).

\subsection{Estimate of the translations between images and co-addition}
\label{subsec:translations}

We estimate the relative motions between the images by cross-correlation. To increase the precision of our estimate of sub-pixel translations between images, we perform a bilinear interpolation on an HR grid at a resolution equal to three times the over-sampling factor ($3 \times f$).
We define the first image as the reference image $\Ib_{\rm ref}$ and we search, in an exhaustive manner, the shift that maximizes the cross-correlation factor, $\rm Cor$, between the reference image and the current translated one, $\Ib$.  
The cross-correlation is computed as:
\begin{equation}
\label{eq:crosscorelation}
\rm Cor(\Ib,  \Ib_{\rm ref})= \frac{1}{\sigma _{\Ib_{\rm ref}} \, \sigma _{\Ib}} \sum _{i_{\rm r}, \, j_{\rm r}} \sum _{i, \, j} \left( \Ib _{i, \, j} - \langle \Ib \rangle \right) \left( \Ib _{\rm ref, \, i, \, j} - \langle \Ib_{\rm ref} \rangle \right) 
\end{equation}
In Eq.~\ref{eq:crosscorelation}, $\sigma _{\Ib_{\rm ref}}$ and $ \sigma _{\Ib}$ are the standard deviations of the matrices $\Ib_{\rm ref}$ and $\Ib$, respectively. $\langle \Ib_{\rm ref} \rangle$ and $\langle \Ib \rangle$ are the averages of $\Ib_{\rm ref}$ and $\Ib$, respectively.

\begin{figure}
  \begin{center}
    \includegraphics[width=0.49\textwidth]{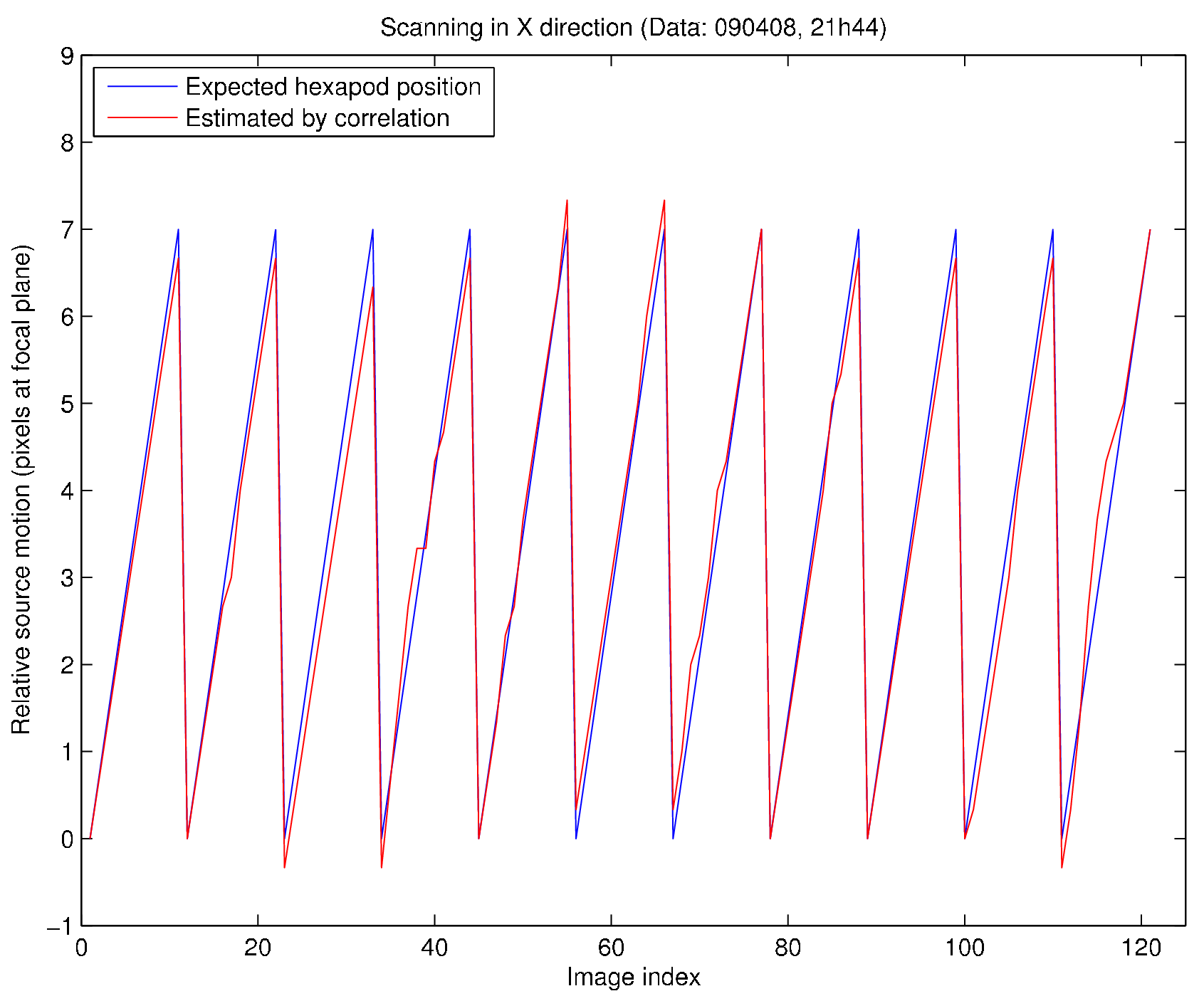}
	\includegraphics[width=0.49\textwidth]{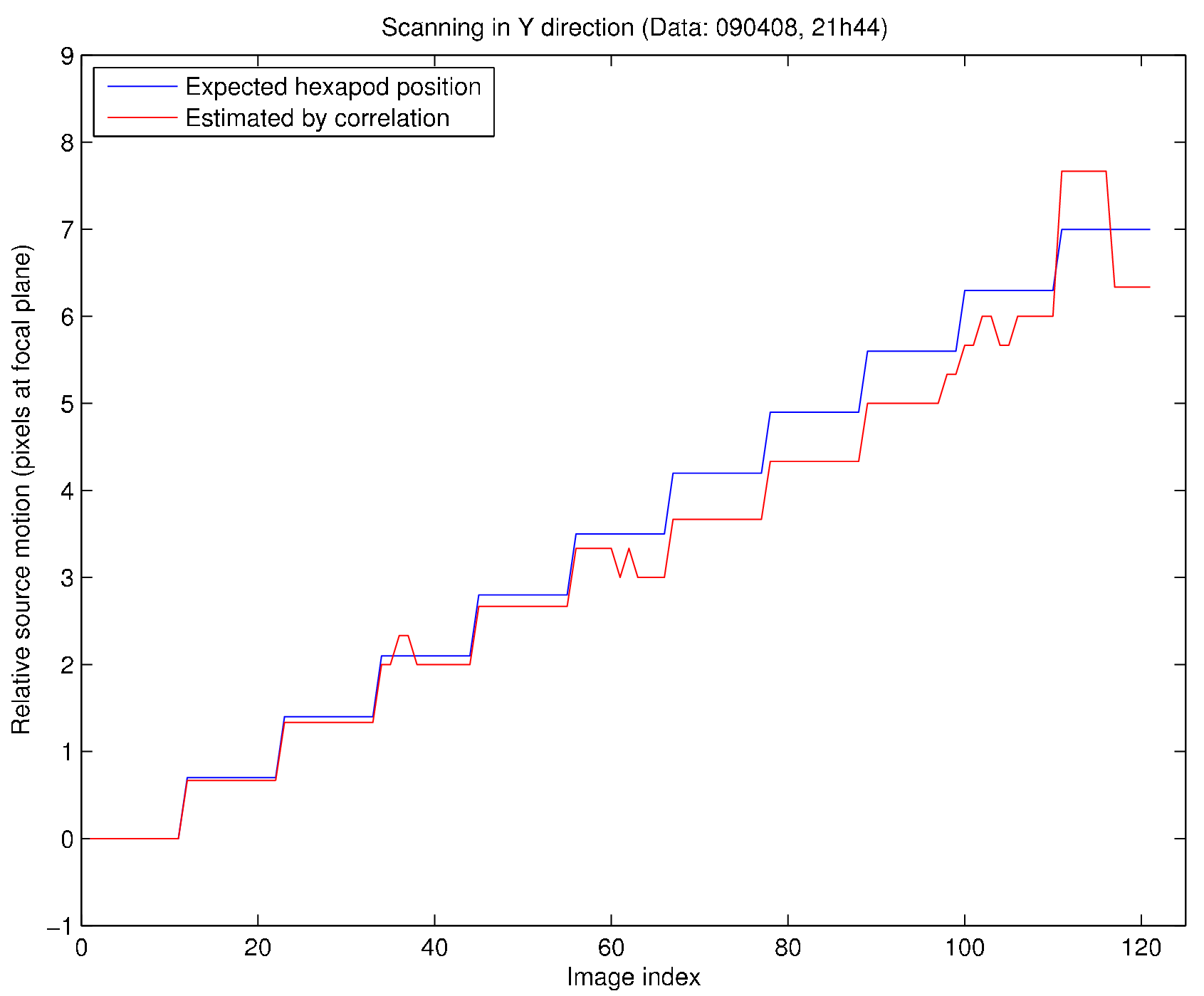}
    \caption[Expected vs. Estimated displacements of the source]{Expected vs. Estimated displacements of the source in the X and Y directions during the microscanning.  A slight discrepancy is seen between the estimated moves on the image and the expected position of the source for the $Y$-direction because, in practice, to control the absolute position of the hexapod is difficult. For the $X$-direction, the agreement is excellent. (Center of the FoV, FM2 data).}
    \label{fig_Deplacement_source}
  \end{center}
\end{figure}

Then, these motions are compared with the expected values measured with the hexapod. 
Fig.~\ref{fig_Deplacement_source} shows an example of the comparison between the read position of the hexapod (expected position in blue) and the position estimated by cross-correlation (in red).
The agreement between expected and estimated positions for the displacement in the $X$-direction is very good. However, a discrepancy is seen for the $Y$-direction. The hexapod did not move at the expected positions in this direction. This issue is likely to be due to the control system of the hexapod support, which is not optimally designed to perform translations along the $X$ and $Y$ axis.  

The mismatch between estimated and expected relative motions between images introduces errors in the reconstruction method if one uses the expected positions. Therefore, the estimated motions by cross-correlation are used to reconstruct the high resolution image. Assuming that the translations $\Tb_k$ are known, we co-add low-resolution images $\yb_k$ onto an over-resolved grid. Thus, a HR image $\zb$ is created, with a number of pixels $f^2$  times larger than the LR image, where $f$ is the \textit{over-sampling factor}.  However, this image is still blurred because of the response of the detector. The slight discrepancy between expected and estimated positions $\Tb_k$ represents a lack of information, in particular in the $Y$-direction. Some pixels of the HR image $\zb$ are not present in the data $\yb_k$.

Ideally, the over-sampling factor may be chosen up to a value of $f=10$, but because of this lack of information, this choice introduces some slight artefacts in the reconstructed images. We chose a value of $f=7$, which appeared to be a good trade-off between the minimization of artefacts and the gain in resolution.

\subsection{Deconvolution}
\label{subsec:deconvolution}


$\zb$ is the convolution of the HR image $\xb$ that we want to calculate by the impulse-response (PSF) of the detector:
\begin{equation}
\label{eq:HRimage}
\zb = \Rb \xb + \nb
\end{equation}
One cannot apply the classical methods of deconvolution because part of the data is not observed (due to uncertainties on the translations, the source is not exactly at its expected position on the fine grid image, see sect.~\ref{subsec:translations}). 
We take into account the truncation of the data by introducing a matrix $\Ab$ in Eq.~\ref{eq:HRimage}:
\begin{equation}
\label{eq:HRimage_T}
\zb = \Ab \Rb \xb + \nb = \Hb \xb +  \nb
\end{equation}
$ \Ab$ is a mask with pixels value of $1$ where the data exists, and 0 otherwise. The operator $\Hb =  \Ab \Rb $ returns the measurements $\zb$ from the HR image $\xb$.

To solve the deconvolution problem, a Bayesian formalism is used, and the lack of information in the data is supplied by the prior. In practice, the deconvolution consists of minimizing a least-square criterion. The estimate of the value $\hat{\xb}$ is given by the argument of the maximum of the \textit{posterior} probability  function, often called maximum  \textit{a posteriori} (MAP). Since the noise $\nb$ is assumed to be Gaussian, independent, and uniform over the whole image, we use a Gaussian likelihood, and we obtain the following least-squares estimator:
\begin{equation}
\label{eq_reg_LS_crit}
\hat{x}_{\rm MAP} =  {\rm arg \  min} ( \Vert z - \Ab \Rb \xb  \Vert ^{2} + \mu \Vert \Db \xb \Vert ^{2} )
\end{equation}
where $\mu$ is the regularization parameter. This equation defines the regularized least-square estimator, where $\Db$ is the finite differences operator (Laplacian).
The least-square criterion, $Q_{\rm LS}$, to minimize in Eq.~\ref{eq_reg_LS_crit} is quadratic and can be re-written as:
\begin{equation}
Q_{\rm LS}(\xb) = (\zb - \Hb \xb) ^{\rm t} (\zb - \Hb \xb) + \mu \xb ^{\rm t} \Db ^{\rm t} \Db \xb
\end{equation}
The solution is explicit:
\begin{equation}
\hat{x}_{\rm ML} = (\Hb ^{\rm t} \Hb + \mu \Db ^{\rm t} \Db)^{-1} \Hb ^{\rm t} \zb
\end{equation}
The main issue with this explicit solution is that the size of the matrix $\Hb ^{\rm t} \Hb$ can be large and its inversion can be very time-consuming. Since the criterion to minimize is quadratic, it is convex and has a unique minimizor. We use a method called ``gradient descent'' or ``steepest descent'', which consists in starting from whatever value for $\xb$, and then converge to the solution in the direction of the steepest slope (opposite of the gradient's direction). At each step, this algorithm ensures that, after a sufficient number of iterations, we will approach the required minimum value of the criterion. Thus we calculate the gradient of the criterion to minimize:
\begin{equation}
g(\xb) =  \frac{\partial Q_{\rm LS}(\xb) }{\partial \xb}  = 2 \, \Hb ^{\rm t} (\Hb \xb - \zb) + 2 \mu \Db ^{\rm t} \Db \xb
\end{equation}
and we update the solution with the following equation:
\begin{equation}
x ^{k+1} = x ^{k} - \alpha _{\rm opt} (\Hb ^{\rm t} (\Hb x ^{k} - \zb) + \mu \Db ^{\rm t} \Db  x ^{k} ) \ ,
\end{equation}
where $\alpha _{\rm opt}$ is the optimal step to ensure the most efficient descent is the opposite direction of the gradient. This step is calculated by cancelling the gradient with respect to $\alpha$. In our case, we use a variant of the gradient descent, so-called ``conjugate gradient descent'', which consists in having orthogonal directions at each descent steps.

\section{Results: the high-resolution PSF of the MIRI imager}
\label{sec:results}

\begin{figure}
  \begin{center}
  \includegraphics[width=\textwidth]{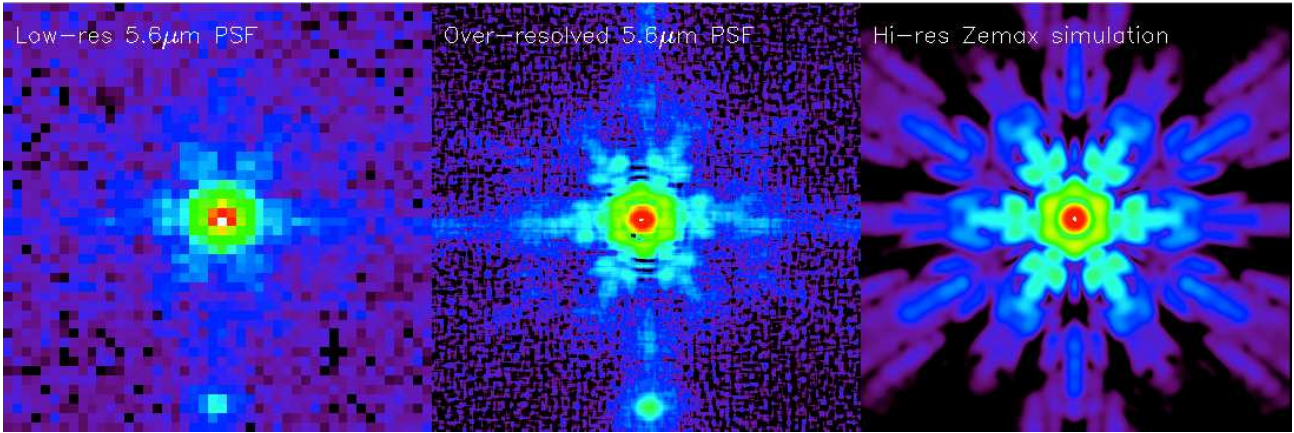}
    \caption{Illustration of the gain in resolution provided by the microscanning and deconvolution method. \textit{Left:} Low-resolution (native) $5.6\,\mu m$ PSF taken in the center of the field of view \textit{Middle:} Corresponding reconstructed high-resolution PSF. \textit{Right:} High-resolution Zemax simulation, to be compared with the reconstructed, over-resolved image. Note that each individual image corresponds to a $5'' \times 5''$ region ($\sim 45 \times 45$ pixels on the LR images, $318 \times 318$ pixels on the high-res images). The ghost spot below the PSF is due to a reflection in the window of the cryostat. All images are displayed on a logarithmic scale to exhibit the secondary lobes. The same dynamic range was used for the display.}
    \label{fig_PSF_LR_HR_Z}
  \end{center}
\end{figure}

Fig.~\ref{fig_PSF_LR_HR_Z} compares the $5.6\,\mu$m PSF at the native resolution (LR, left image), with the over-resolved PSF obtained with the deconvolution method described above (middle image), and the simulated (Zemax) PSF at high-resolution (HR).  The linearity correction is applied on the data and the over-sampling factor is set to $f=7$.
LR data correspond to $25\,\mu$m pixels on the detector, wheras HR correspond to a pixel size of $3.6\,\mu$m.  One original MIRIM pixel equals to $7^2$ pixels in the over-resolved image. 

The secondary Airy diffraction ring, which is barely visible on LR images, is clearly exhibited on the HR images. The shape of the secondary lobes is resolved on the deconvolved image, and agrees very well with Zemax simulations, as well as the relative intensities of the secondary lobes with respect to the central peak.
On the over-resolved PSF, we note the presence of secondary dark rings in the $Y$ direction, as well as pronounced diffraction spikes in the $X$ and $Y$ directions, on top of the normal JWST diffraction pattern. The dark rings are artefacts caused by the lack of information introduced by the errors of the hexapod motions in the $Y$ direction (see Sect.~\ref{subsec:translations}).
The root cause of the diffraction spikes is under investigation, and may be due to the coating of the mirrors. 
Those diffraction features are only visible at very high signal to noise ratio (typically $> 2000$). In addition, a ghost spot is present below the PSF, due to the reflection in the window of the cryostat.

The PSF images were analysed with a set of IDL routines, complemented with the use of the IRAF software. Different measurements of the FWHM of the PSFs have been performed (Gauss, Airy fits, radial profiles, encircled energy, radius containing 65\% of the total energy). 

\begin{figure}
  \begin{center}
    \includegraphics[height=0.49\textwidth, angle = 90]{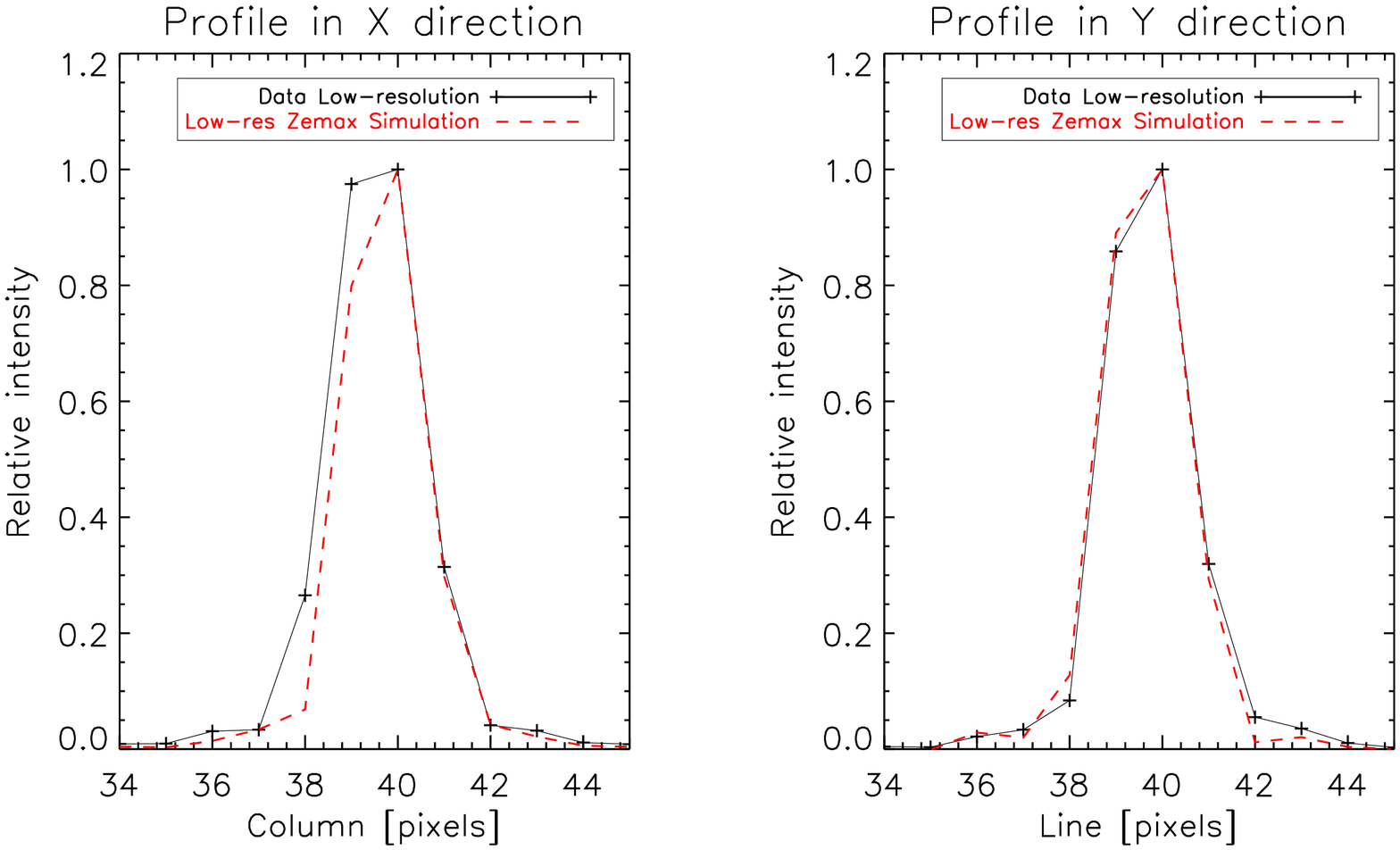}
    \includegraphics[height=0.49\textwidth, angle = 90]{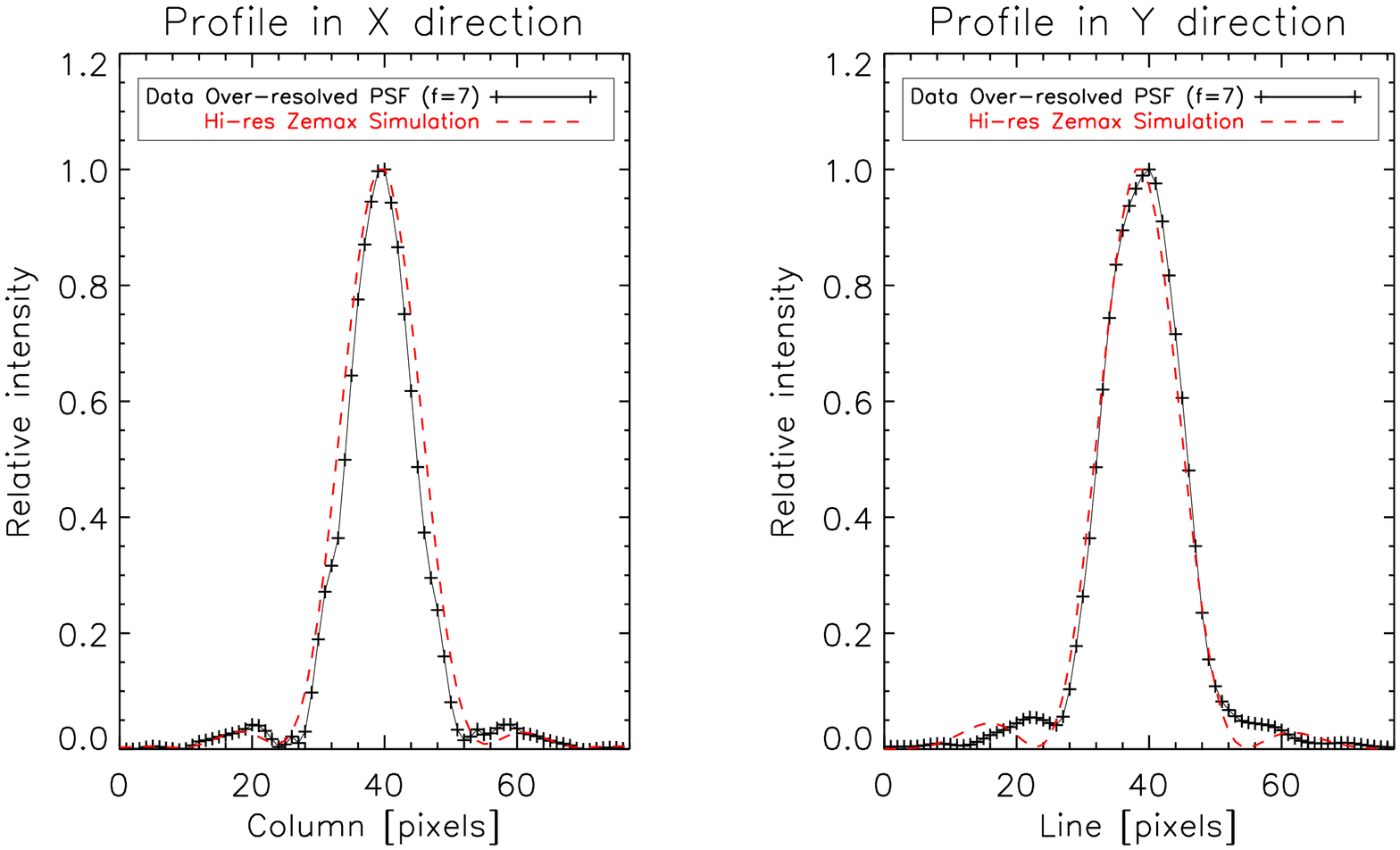}
    \caption{Low (2 first plots on the left ) and over-resolved (2 last plots on the right) PSF profiles in the X and Y directions. For the high-resolution PSF, the sampling is increased by a factor of 7. The data were acquired during the FM4 campaign, in the center of the FoV. We compare the low and high resolution PSF to low and high resolution Zemax simulations, respectively (red dashed lines). The Zemax simulations shown are in the nominal, polychromatic case.}
    \label{fig_PSF_profiles}
  \end{center}
\end{figure}

\subsection{Comparison of the results of the three test campaigns (FM1, FM2 and FM4)}
\label{subsec:test_campaigns}

\begin{table}
\begin{minipage}{\columnwidth}
 \renewcommand{\footnoterule}{}
\def\thefootnote{\alph{footnote}}
 \caption[]{PSF measurements for the data taken during the differents test campaigns at CEA\footnotemark[1]. }
 \centering
\begin{tabular}{cccccc}
\hline
\hline
\multirow{2}*{Campaign} & \multirow{2}*{Method} &  \multicolumn{4}{c}{Data} \\
\cline{3-6}
& &  \multicolumn{2}{c }{Low-Resolution} & \multicolumn{2}{c }{Over-resolved $f=7$}  \\ 
\hline
\multirow{4}*{FM1}  & \multirow{2}*{Airy}  & $X$ & $Y$ & $X$ & $Y$  \\
\cline{3-6}
& & 56.84 & 55.61 & 46.93 & 47.29 \\
\cline{2-6}
& R$_{65\%}$ &  \multicolumn{2}{c}{46.79} &  \multicolumn{2}{c}{37.96}  \\
\cline{2-6}
& E$_{\rm cent}$  &  \multicolumn{2}{c}{0.47} &  \multicolumn{2}{c}{0.51} \\
\hline
\hline
\multirow{4}*{FM2}  & \multirow{2}*{Airy}  & $X$ & $Y$ & $X$ & $Y$  \\
\cline{3-6} 
&    & 53.77 & 51.31 &  45.73 & 44.04  \\
\cline{2-6}
& R$_{65\%}$ &  \multicolumn{2}{c}{42.42} &  \multicolumn{2}{c}{33.10}  \\
\cline{2-6}
& E$_{\rm cent}$  &  \multicolumn{2}{c}{0.50} &  \multicolumn{2}{c}{0.56} \\
\hline
\hline
\multirow{4}*{FM4}  & \multirow{2}*{Airy}  & $X$ & $Y$ & $X$ & $Y$  \\
\cline{3-6}
&    & 54.21  & 50.50  &  42.26  & 43.61   \\
\cline{2-6}
& R$_{65\%}$ &  \multicolumn{2}{c}{41.90} &  \multicolumn{2}{c}{32.64}  \\
\cline{2-6}
& E$_{\rm cent}$  &  \multicolumn{2}{c}{0.51} &  \multicolumn{2}{c}{0.58} \\
\hline
\hline
\end{tabular}
\footnotetext[1]{Results are indicated for the center of the MIRIM field of view. For the HR image, the sampling is increased by a factor of 7 with respect to a native MIRIM image, i.e. one original MIRIM pixel equals to $7^2$ pixels in the over-resolved image. Three different measurements are indicated: the FWHM of the 2D Airy --Bessel function-- fit ($X$ and $Y$ directions in $\mu$m), the radius containing 65\% of the energy in $\mu$m, and the fraction of the total energy contained in the central lobe, normalized to 5''. The statistical $1\, \sigma$ uncertainties on the fitted FWHM values are of the order of $0.04\, \mu$m.}
\label{tab_results_data}
\end{minipage}
\end{table}

\begin{table}
\begin{minipage}{\columnwidth}
 \renewcommand{\footnoterule}{}
\def\thefootnote{\alph{footnote}}
 \caption[]{PSF measurents for the Zemax simulations\footnotemark[1]. }
 \centering
\begin{tabular}{ccccccccc}
\hline
\hline
\multirow{2}*{Setup} & \multirow{2}*{Method}  & \multicolumn{4}{c}{Simulations} \\
\cline{3-6}
&   & \multicolumn{2}{c}{Zemax Low-R}& \multicolumn{2}{c}{Zemax Hi-res}\\ 
\hline
\multirow{4}*{nominal}  & \multirow{2}*{Airy}  & $X$ & $Y$ & $X$ & $Y$  \\
\cline{3-6}
& &  48.62 & 50.45 & 42.55 & 42.47 \\
\cline{2-6}
& R$_{65\%}$ &  \multicolumn{2}{c}{39.14} &  \multicolumn{2}{c}{28.59}  \\
\cline{2-6}
& E$_{\rm cent}$  &  \multicolumn{2}{c}{0.64} &  \multicolumn{2}{c}{0.72} \\
\hline
\hline
\multirow{4}*{tolerance}  & \multirow{2}*{Airy}  & $X$ & $Y$ & $X$ & $Y$  \\
\cline{3-6}
&    & 49.78 & 51.96 & 44.26 & 44.89 \\
\cline{2-6}
& R$_{65\%}$ &  \multicolumn{2}{c}{41.26} &  \multicolumn{2}{c}{30.12} \\
\cline{2-6}
& E$_{\rm cent}$ &  \multicolumn{2}{c}{0.61} &  \multicolumn{2}{c}{0.68}\\
\hline
\hline
\end{tabular}
\footnotetext[1]{Results are indicated for two setups of Zemax simulations: a nominal case, and a case where mechanical tolerances were applied in the Monte-Carlos (see text for details).}
\label{tab_results_Zemax}
\end{minipage}
\end{table}

The results of the PSFs measurements in the center of the MIRIM FoV are gathered in Tables~\ref{tab_results_data} and \ref{tab_results_Zemax}. Table~\ref{tab_results_data} shows the PSF characteristics  for the three test campaigns carried out at CEA. Table~\ref{tab_results_Zemax} indicates the results of Zemax simulations in two cases:  a \textit{nominal} case, and a case where mechanical \textit{tolerances} are included. In the latter, we select the most extreme deviation from the nominal case observed in the Monte Carlo draws. The Zemax simulations are polychromatic, i.e. they include the width of the $5.6\,\mu$m filter bandpass. 

We first note that the low-resolution PSFs are systematically wider than the over-resolved ones, which is due to the fact that the native $5.6\,\mu$m PSF is not Nyquist-sampled. This illustrates that the microscanning technique, and associated deconvolution method, are crucial to verify precisely the optical quality of the imager at this wavelength.
Examples of PSF profiles (cut in the $X$ and $Y$ directions) are given in Fig.~\ref{fig_PSF_profiles}. 
The comparison of the LR and HR profiles illustrates the gain in resolution. 
The better sampling of the PSF helps in reconstructing more accurately the main peak of the PSF. The secondary diffraction lobes are also resolved. 
The observed PSF are overall in good agreement with nominal Zemax simulations. The cores of the PSFs are also well fitted by Bessel functions (labeled as Airy fit in the Tables), although we note that the positions and amplitudes of the secondary lobes do not match exactly with the Airy fits (neither with the simulations).

\begin{figure}
  \begin{center}
  \includegraphics[width=\textwidth]{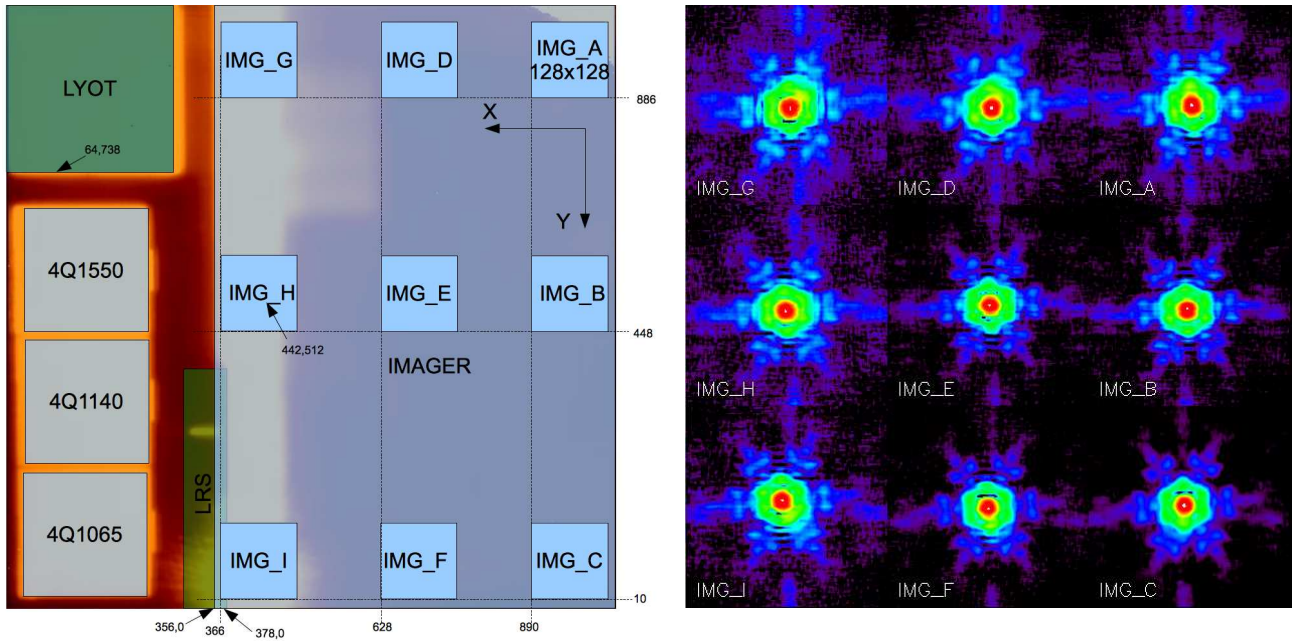}
    \caption{\textit{Left:} Definition of the subarrays used during cold performance tests at CEA during the FM4 campaign. \textit{Right:} Reconstructed high-resolution PSF at $5.6\,\mu m$ for each of the 9 microscans performed in imager field of view (subarrays A to I). Note that each individual image corresponds to a zoom on the PSF equivalent to $\sim 28 \times 28$ pixels on the detector ($200 \times 200$ pixels on the high-res image).}
    \label{fig_PSF_subarrays}
  \end{center}
\end{figure}

\begin{description}
\item[FM1 test campaign exhibits a PSF and vigneting issue.]
An inspection of the FM1 results shows that the FWHM of the HR PSFs are systematically larger than the measured FWHM on the HR Zemax PSF. The discrepancy is about $+8$~$\mu$m. This mismatch is out of the specifications of the instrument in terms of wavefront error and encircled energy.
This issue was also revealed independently by the measurement of the MIRIM FoV, which showed a vignetting issue. 
Motivated by these tests, a careful inspection and a new metrology analysis of the instrument was carried out at CEA in cooperation with IAS, Orsay. These investigations concluded to an interference between the structure of the Three-Mirror-Anastigmatic (TMA) objective and the M4 mirror. This design problem caused a tilt of the M4 mirror with respect to its nominal positioning. It was observed that the rear face of mirror M4 was not flat and was tilted by 0.33 degree with respect to the fixation pads.
The M4 mirror has been found damaged, with some bending of its mounting flexible pad. This tilt resulted in a degradation of the nominal PSF properties and the FoV area.

\item[FM2 campaign: requirements met after correction for the tilt of the M4 mirror.]
The cause of the interference between the TMA structure and the M4 mirror was removed manually. The damaged M4 mirror was replaced by the Flight Spare model. 
A second test campaign (FM2) was launched, and we performed the same microscanning analysis to check the optical quality of the PSF. We note that for this campaign (as well as for the FM1), the filter wheel assemble was not available. It was replaced by a ``cyclop tool'' which allows us to mount the $5.6\,\mu$m filter. The results show that the correction of the M4 tilt improved the optical quality. 
The measured FWHM with Airy fits on the over-resolved PSFs are of the order of $44-48 \, \mu$m for both $X$- and $Y$- directions, and over the whole FoV, which is about 
$2-4$~$\mu$m wider than the HR simulated PSF. 
This corresponds to FWHM of $0.19-0.21$~arcsec. In addition,  more than $56\,$\% of the total encircled energy within a 5'' radius is contained in the first dark Airy ring (see Sect.~\ref{subsec:EE}). This is compliant with the OBA-1004 requirement\footnote{OBA-1004: MIRI Imager data shall include $> 56\,$\% of the Encircled Energy of the image of a point source at 
wavelengths longward of $5.6\,\mu$m within the angular diameter of the first dark ring, as determined from the 
PSF generated using the segmented telescope model in the IRD. Note: The Encircled Energy is defined 
relative to the energy within a radius of 5 arcsec equivalent projected onto the sky. }.

\item[FM4 campaign: final verification of the optical quality with the FWA.]
For the FM4 test campaign, the Filter Wheel Assembly was available and integrated into MIRIM, thus providing PSF measurements at wavelengths $5.6 - 25.5\,\mu$m. 
The measurements show a very good agreement between the reconstructed HR PSF and the HR nominal Zemax simulations. Fig.~\ref{fig_PSF_subarrays} shows the over-resolved PSF in 9 regions of the MIRIM FoV (subarrays A to I). We found that the OBA-1004 requirement is met in all 9 regions (see Sect.~\ref{subsec:EE}). At longer wavelengths ($7.7-25.5\,\mu$m), $57-68\,$\% of the encircled energy is contained within the first dark ring radius, which confirms that MIRIM is compliant with the optical quality requirements in terms of encircled energy (see Ronayette et al. 2010\cite{Ronayette2010}). 
However, we note that both $R_{65\%}$\footnote{$R_{65\%}$ is the radius containing 65\% of the total energy within a 5 arcsec radius}, and $E_{\rm cent}$\footnote{$E_{\rm cent}$ is the fraction of the total energy contained in the central lobe, normalized to 5 arcsec.} indicate that the simulated PSF with tolerances (in the worst case) is of better quality than the reconstructed high-resolution PSF (Tables~\ref{tab_results_data} and \ref{tab_results_Zemax}). We are still investigating to understand why this is the case.
The HR technique also allows us to identify some deformations of the secondary hexagonal lobe across the FoV, as well as a displacement of the diffraction spike.

\end{description}

\subsection{Encircled energy}
\label{subsec:EE}

\begin{figure}
  \begin{center}
    \includegraphics[angle=90,width=0.9\textwidth]{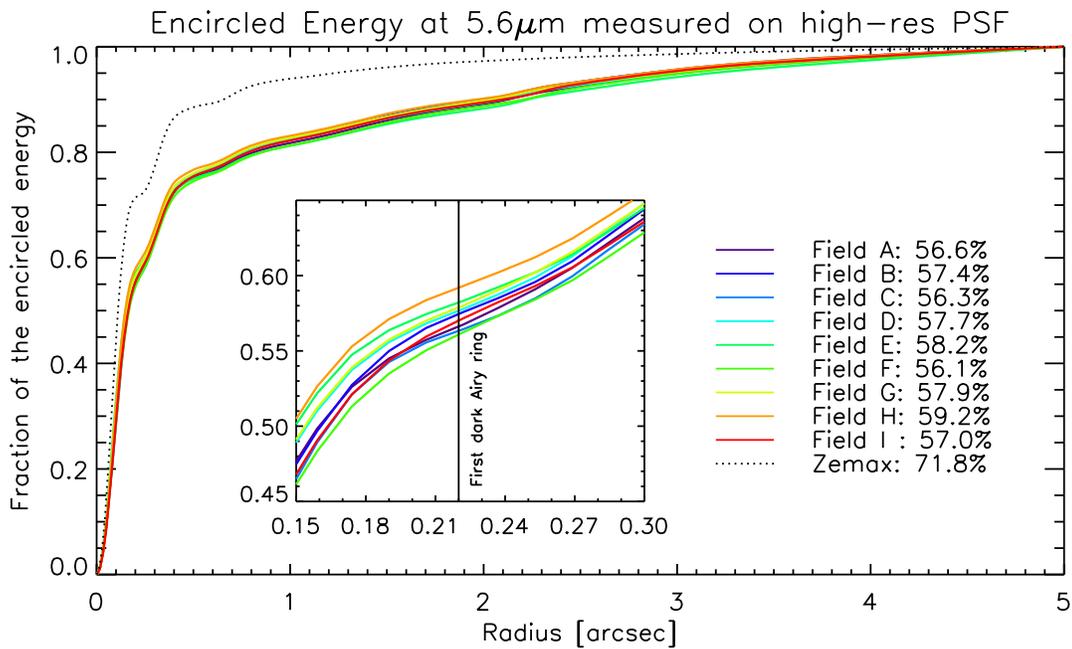}
    \caption{Encircled energy of the MIRI high-resolution (HR) PSF  as a function of radius in arcsec (1 pixel = 0.11''). The solid lines are the results for the nine over-resolved PSF in different areas of the imager field of view. The dashed line shows a nominal PSF simulated with the Zemax software. The inflection point corresponds to the first dark ring of the Airy disk of the PSF. The percentages indicate the fraction of the  encircled energy within the first dark Airy ring radius (situated at the radius $47.4\,\mu$m).}
    \label{fig:EE}
  \end{center}
\end{figure}

This section discusses further the optical quality of the MIRIM PSF in terms of  encircled energy, since the OBA requirements are expressed in these terms.
The encircled energy $E(r)$ of a radially symmetric point spread function $\phi(r)$ can be defined as the integral
\begin{equation}
E(r) = 2 \pi \int_{0}^{r} \phi (r') d r'
\end{equation}
For the purpose of comparison to the requirement document, we normalised $E(r)$ to 1 at $r =5~$arcseconds. However, our microscanning data acquisition was done for a sub-array of $\approx 7'' \times 7''$ during the FM1 and FM2 campaigns to speed up the data acquisition. To normalize the encircled energy at a radius of 5'', we used the background of a full-frame PSF image. For the FM4 test campaign, the sub-arrays are $\approx 14'' \times 14''$, which allows us to normalize the encircled energy directly.

Fig.~\ref{fig:EE} shows the encircled energy of HR reconstructed PSFs in the different regions of the MIRIM FoV,  as a function of radius in arcsec (1 pixel = 0.11''). The dotted line shows the encircled energy of the HR Zemax PSF, computed in the nominal case, and in the center of the FoV.  The first point of inflection in $E(r)$, i.e the value of $r$ for which $d^2E/dr^2 = 0$, corresponds to the diameter of the first dark Airy ring. We find a fraction of the encircled energy of $56.1-59.2\,$\% inside the central lobe over the whole FoV, which is compliant with the OBA-1004 requirement.

We note that the measured variation of encircled energy with radial distance from a point image is very sensitive to the flux levels at large radial distances. The condition $dE/dr \geq 0$ for all values of $r$ must be satisfied for the measurement to make sense. Here, this is the case because the  background values at large radial distances are high. The telescope simulator is at ambient temperatures, indeed producing a high thermal background.

\section{Summary and concluding remarks}
\label{sec:conclusion}

In this paper we have presented the first detailed analysis of the JWST Mid-InfraRed IMager (MIRIM) Point Spread Function (PSF) at 5.6$\, \mu$m, from three  cryogenic test campaigns performed at CEA. We used a high-resolution (HR) microscanning technique  to verify the MIRIM optical quality and characterize the 5.6$\, \mu$m PSF with an unprecedent accuracy. The microscanning consists in a fine, sub-pixel scanning of a point source on the focal plane. A deconvolution algorithm is used to reconstruct over-resolved PSFs (by a factor of 7 times the native resolution). We summarize here our main results:

\begin{enumerate}
\item We characterized the response of the detector, which is non-linear  above $20\,000$~ADU. The correction for the non-linearity improves the quality of the MIRIM PSFs by a factor of $\sim 5\,$\%. Note that the detector and electronics used for these tests were not flight model hardware. A less severe non-linear response is expected on the FM detectors.

\item The microscanning test and associated deconvolution provide a significant improvement for the characterization of the PSF. This test allows to resolve the diffraction pattern of MIRIM and to measure accurately the width of the PSF for the flight model. The FWHM and the shape of the PSF secondary lobes is as expected from HR Zemax simulations.

\item The FWHM of the HR reconstructed images taken during the FM1 test campaign were  $4-6\,\mu$m wider than HR  Zemax simulations, and the encircled energy was out-of-specifications. The root cause was a tilt of the M4 mirror. After correction, we show that the observed HR PSF is in excellent agreement with simulations. The FWHM is 0.19'' - 0.20'', and $56.1-59.2$\% of the total encircled energy within a 5'' radius is inside the central lobe, over the whole field of view, which is within specifications. At longer wavelengths ($7.7-25.5\,\mu$m), $57-68\,$\% of the encircled energy is contained within the first dark ring radius, which confirms that MIRIM is compliant with the optical quality requirements.

\item Slight differences between the simulated PSF patterns and the MIRIM HR reconstructed data are exhibited. Some of these features (dark arcs) are associated with the  limits of the deconvolution method itself, in particular due to errors in the relative positions of the images and the associated lack of information. The other features (deformation of the secondary lobes and diffraction spikes) require further investigation to understand their root cause.
\end{enumerate}

The microscanning technique and deconvolution algorithm described in this paper have been particularly powerful to verify the optical performance of MIRIM and identify defects in the instrument. This high-precision characterization of the PSF at the shortest operating wavelength will be of importance for the photometric calibration of MIRI, but also for possible scientific applications, like point source extraction or map-making, where a precise knowledge of the PSF is required. This technique may also be applied for the MIRI Medium Resolution Spectrometer (MRS) integral field unit. In addition, we note that the dithering mirror implemented on the JWST may be used to perform over-resolution, following the methods described here to deconvolve the images. This may push even further the capabilities of the JWST in terms of angular resolution.


\bibliography{microscan_miri.bbl}
\bibliographystyle{spiebib}   

\end{document}

%% file: abrmath.tex
\def\AND{\text{and\:}}           
\def\OR{\text{or\:}}

\newsavebox{\fminibox}
\newlength{\fminilength}

  \def\+{^\dagger}

\def\nequiv{\not\kern-.05em\equiv}
\def\egal{\kern-.5em=\kern-.5em}        
\def\propt{\kern-.2em\propto\kern-.2em} 

\def\intdouble{\int\kern-0.3em\int}
\def\inttriple{\int\kern-0.3em\int\kern-0.3em\int}

\def\rond#1{\overset{\kern-0.33em~_\circ}{#1}}
\def\rondit[#1]#2{\overset{\kern#1~_\circ}{#2}}
